\documentclass[a4paper,conference,final]{IEEEtran}
\usepackage{cite}
\usepackage[binary-units=true]{siunitx}
\DeclareSIUnit \dBm {dBm}
\DeclareSIUnit \dB {dB}
\DeclareSIUnit \dBi {dBi}
\DeclareSIUnit \Kbps {Kbps}
\DeclareSIUnit \Mbps {Mbps}
\DeclareSIUnit \Gbps {Gbps}
\DeclareSIUnit \kBps {kBps}
\DeclareSIUnit \MBps {MBps}
\DeclareSIUnit \GBps {GBps}
\DeclareSIUnit \GHz {GHz}
\usepackage{graphicx}
\usepackage{amsfonts}
\usepackage{multirow}
\usepackage{amsmath}
\usepackage{nicefrac}
\usepackage{tabularx}
\usepackage{enumerate}
\usepackage{tikz}
\usetikzlibrary{arrows,automata}

\usepackage[caption=false,font=footnotesize]{subfig}
\usepackage[hyphens]{url}
\usepackage{algorithm}
\usepackage{algpseudocode}

\newcolumntype{P}[1]{>{\raggedright\arraybackslash}p{#1}}
\newcolumntype{M}[1]{>{\centering\arraybackslash}m{#1}}
\newcolumntype{C}{ >{\centering\arraybackslash} m{2.3cm} }
\begin{document}
\title{Operating ITS-G5 DSRC over Unlicensed Bands: A City-Scale Performance Evaluation}
\author{\IEEEauthorblockN{\thanks{I.M. and A.T. developed the communication infrastructure, recorded and analyzed the data. R.J.P. assisted with editing.}Ioannis Mavromatis\IEEEauthorrefmark{1}, Andrea Tassi\IEEEauthorrefmark{1} and Robert J. Piechocki\IEEEauthorrefmark{1}\IEEEauthorrefmark{2}}
\IEEEauthorblockA{\IEEEauthorrefmark{1}Department of Electric and Electronic Engineering, University of Bristol, UK}
\IEEEauthorblockA{\IEEEauthorrefmark{2}The Alan Turing Institute, London, NW1 2DB, UK}
Emails: \{Ioan.Mavromatis, A.Tassi, R.J.Piechocki\}@bristol.ac.uk}

\maketitle

\begin{abstract}
Future Connected and Autonomous Vehicles (CAVs) will be equipped with a large set of sensors. The large amount of generated sensor data is expected to be exchanged with other CAVs and the road-side infrastructure. Both in Europe and the US, Dedicated Short Range Communications (DSRC) systems, based on the IEEE 802.11p Physical Layer, are key enabler for the communication among vehicles. Given the expected market penetration of connected vehicles, the licensed band of $\SI{75}{\mega\hertz}$, dedicated to DSRC communications, is expected to become increasingly congested. In this paper, we investigate the performance of a vehicular communication system, operated over the unlicensed bands $\SI{2.4}{\giga\hertz}$-$\SI{2.5}{\giga\hertz}$ and $\SI{5.725}{\giga\hertz}$-$\SI{5.875}{\giga\hertz}$. Our experimental evaluation was carried out in a testing track in the centre of Bristol, UK and our system is a full-stack ETSI ITS-G5 implementation. Our performance investigation compares key communication metrics (e.g., packet delivery rate, received signal strength indicator) measured by operating our system over the licensed DSRC an the considered unlicensed bands. In particular, when operated over the $\SI{2.4}{\giga\hertz}$-$\SI{2.5}{\giga\hertz}$ band, our system achieves comparable performance to the case when the DSRC band is used. On the other hand, as soon as the system, is operated over the $\SI{5.725}{\giga\hertz}$-$\SI{5.875}{\giga\hertz}$ band, the packet delivery rate is $30\%$ smaller compared to the case when the DSRC band is employed. These findings prove that operating our system over unlicensed ISM bands is a viable option. During our experimental evaluation, we recorded all the generated network interactions and the complete data set has been publicly available.
\end{abstract}


\vspace{-2mm}
\section{Introduction}
By the end of 2020, recent forecasts estimate that fifty billion devices will require internet connectivity to operate. Among these devices, around ten million vehicles equipped with multiple communication systems and autonomous capabilities are expected to be rolled out into the global market~\cite{cisco}. In particular, there is global consensus around the fact that future automotive services for Connected and Autonomous Vehicles (CAVs) are expected to rely heavily on reliable broadband connectivity on the move~\cite{8281108}. This is confirmed by the National Highway Traffic Safety Administration (U.S. Department of Transportation) and the European Commission's Connected-Intelligent Transportation System (C-ITS) initiative~\cite{NHTSA,C_ITS}. In particular, the latter, along with the 5G-PPP Partnership, argues how reliable Vehicle-to-Everything (V2X) connectivity will empower future CAVs to deliver advanced automotive services, such as See-Through, Automated Overtake and High-Density Platooning~\cite{5G-PPP,10.4108/eai.20-3-2018.154368}.

As identified by the C-ITS initiative, future CAVs are expected to be equipped with over 200 sensors. These sensors will generate a potentially substantial data stream to be shared with surrounding CAVs~\cite{RR0.3}. Given the expected market penetration of CAVs and their need of reliable communications links, Dedicated Short Range Communications (DSRC) systems based on the IEEE 802.11p Physical Layer~\cite{5771027} are expected to become more and more congested.
This directly follows from the fact that users will access the channel on a contention-based way, similar to IEEE 802.11a~\cite{6737584}, sharing a very narrow licensed frequency band (namely, $\SI{5.85}{\giga\hertz}$-$\SI{5.925}{\giga\hertz}$).
What is more, there are plans for operating Cellular-V2X (C-V2X) systems on at least fractions of the band dedicated to DSRC systems, which will impact on the network resources available for ITS-G5 systems~\cite{8374081}.

In an effort of addressing the pressing concern of the expected reduction of network resource over the DSRC licensed band, we investigate the possibility of using unlicensed bands in conjunction with the licensed DSRC band for supporting V2X communications. In particular, we establish a performance evaluation of a full-stack implementation of the ETSI's ITS-G5 DSRC system over Industrial, Scientific, and Medical (ISM) bands. In our experiments, we considered the ISM bands $\SI{2.4}{\giga\hertz}$-$\SI{2.5}{\giga\hertz}$ (hereafter referred to as ``ISM-2.4'') and $\SI{5.725}{\giga\hertz}$-$\SI{5.875}{\giga\hertz}$ (hereafter referred to as ``ISM-5''). In order to do so, we deployed multiple ITS-G5 Road-Side Units (RSUs) along a $\SI{5}{\kilo\metre}$ long testing track located in the centre of Bristol, UK. We equipped two vehicles with ITS-G5 On-Board Units (OBUs) that has been driven around the testing track for $\SI{4}{\hour}$ per day, over a period of time of four days ($\SI{16}{\hour}$ in total). During our experiments, every network interaction between the OBUs and RSUs has been recorded along with the positioning information of both transmitter and receiver. The resulting data set is the first to record the whole set of ITS-G5 network interactions over a large-scale environment across both licensed and unlicensed bands.

The coexistence of IEEE 802.11p-based systems has been extensively investigated for what concerns the possibility of operating DSRC systems over unlicensed DVB-T2 bands~\cite{7177281}. On the other hand,~\cite{6737584,8275636} investigated the coexistence of IEEE 802.11a/n/ac technologies as secondary systems over the band dedicated for DSRC applications. To the best of knowledge, no large-scale performance investigation across both the ISM-5 and ISM-2.4 bands exists. In particular, the original contributions of this paper are summarized as follows:
\begin{itemize}
    \item We compare the network performance in terms of Packet Delivery Rate (PDR) and Received Signal Strength Indicator (RSSI) across ISM bands commonly used for operating WiFi networks. In particular, we observe that, on average, when the system is operated over the considered ISM-5 bands, the PDR is no more than $30\%$ lower compared to the correspondent case when the system is operated over the licensed DSRC band, while our system achieved almost comparable performance in the ISM-2.4 band.
    \item We released the entire data set of network interactions to enable future network comparisons. In particular, the structure of our data set two-folded: (i) a Comma-Separated Values (CSV) database to provide agile access to positioning information and transmission metrics and, (ii) A set of Packet Capture (PCAP) traces enabling to extract further system metrics (not previously included in the CSV database).
\end{itemize}

The remainder of the paper is organized as follows. Section~\ref{sec.2} presents the considered experimental setup. Section~\ref{sec.3} presents the structure of our car trials. Our findings are discussed in Section~\ref{sec.4}. In Section~\ref{sec.5}, we draw our conclusions.

\section{System Description}\label{sec.2}

\begin{figure*}[t]
\begin{minipage}{.46\linewidth}
\centering
\subfloat[Different components of our prototyped OBU device.]{\label{fig:close}\includegraphics[width=0.98\linewidth]{{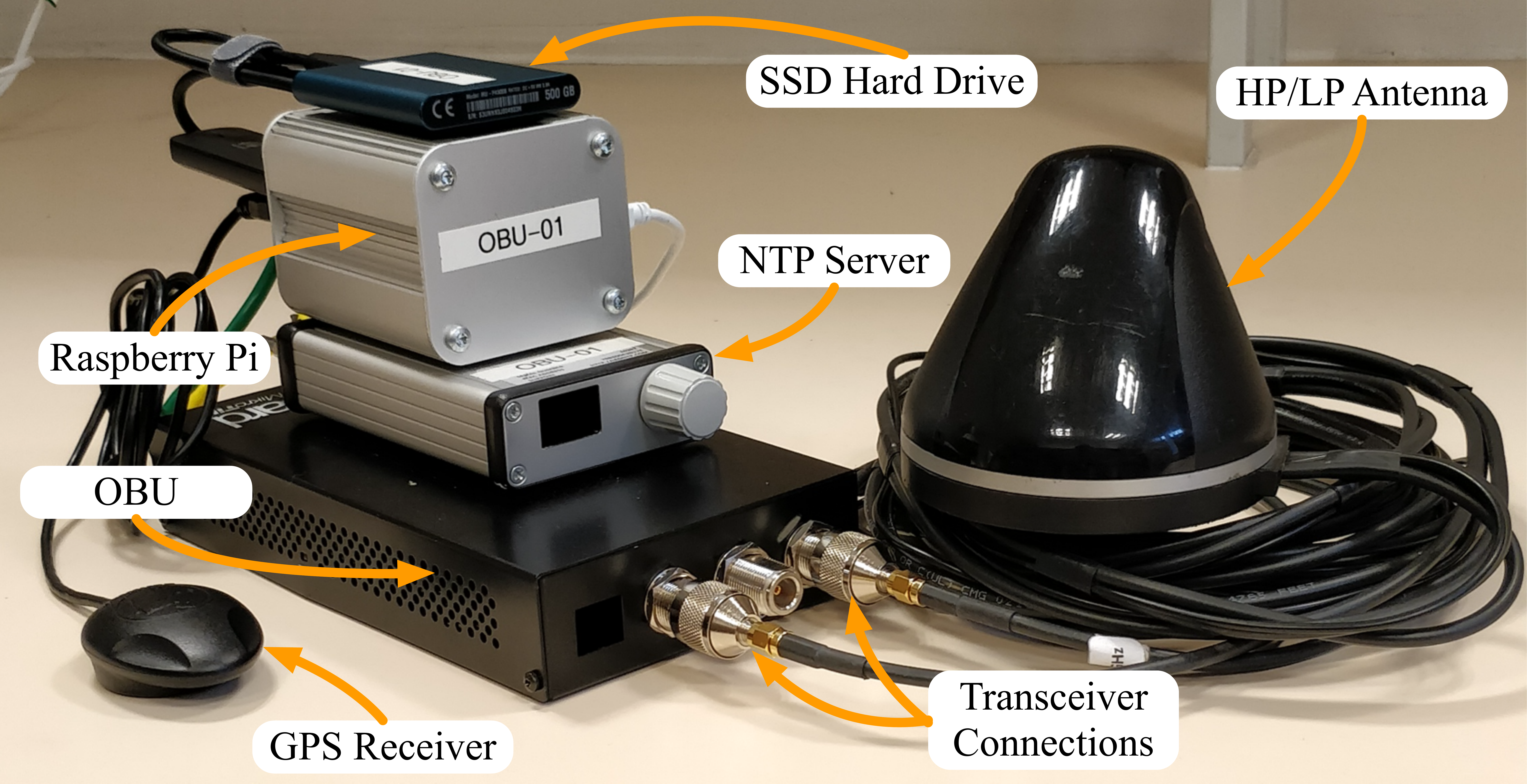}}}
\end{minipage}
\begin{minipage}{.54\linewidth}
\centering
\subfloat[IEEE 802.11p/DSRC OBU system setup.]{\label{fig:boot}\includegraphics[width=0.98\linewidth]{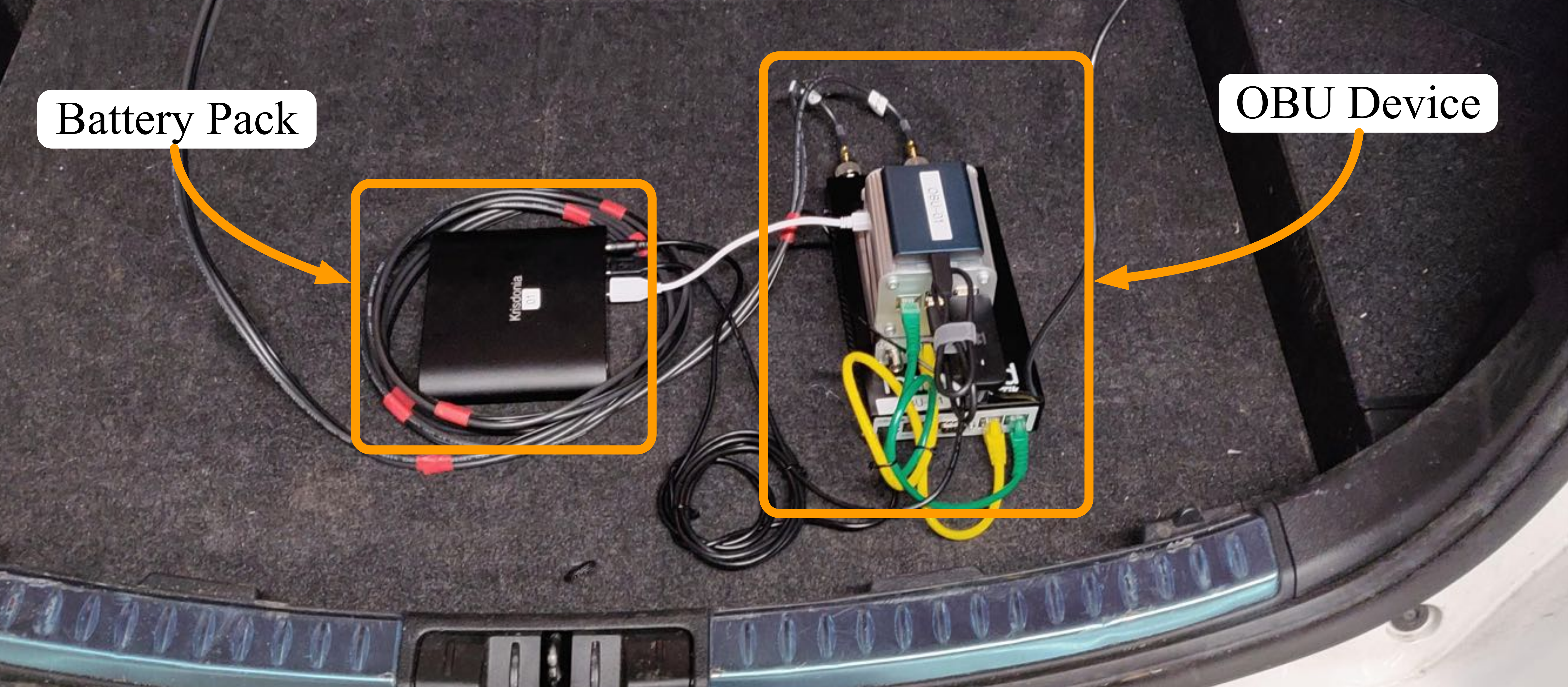}}
\end{minipage}\par\medskip
\centering
\begin{minipage}{.20\linewidth}
\centering
\subfloat[Prototyped RSU units.]{\label{fig:rsu}\includegraphics[width=0.98\linewidth]{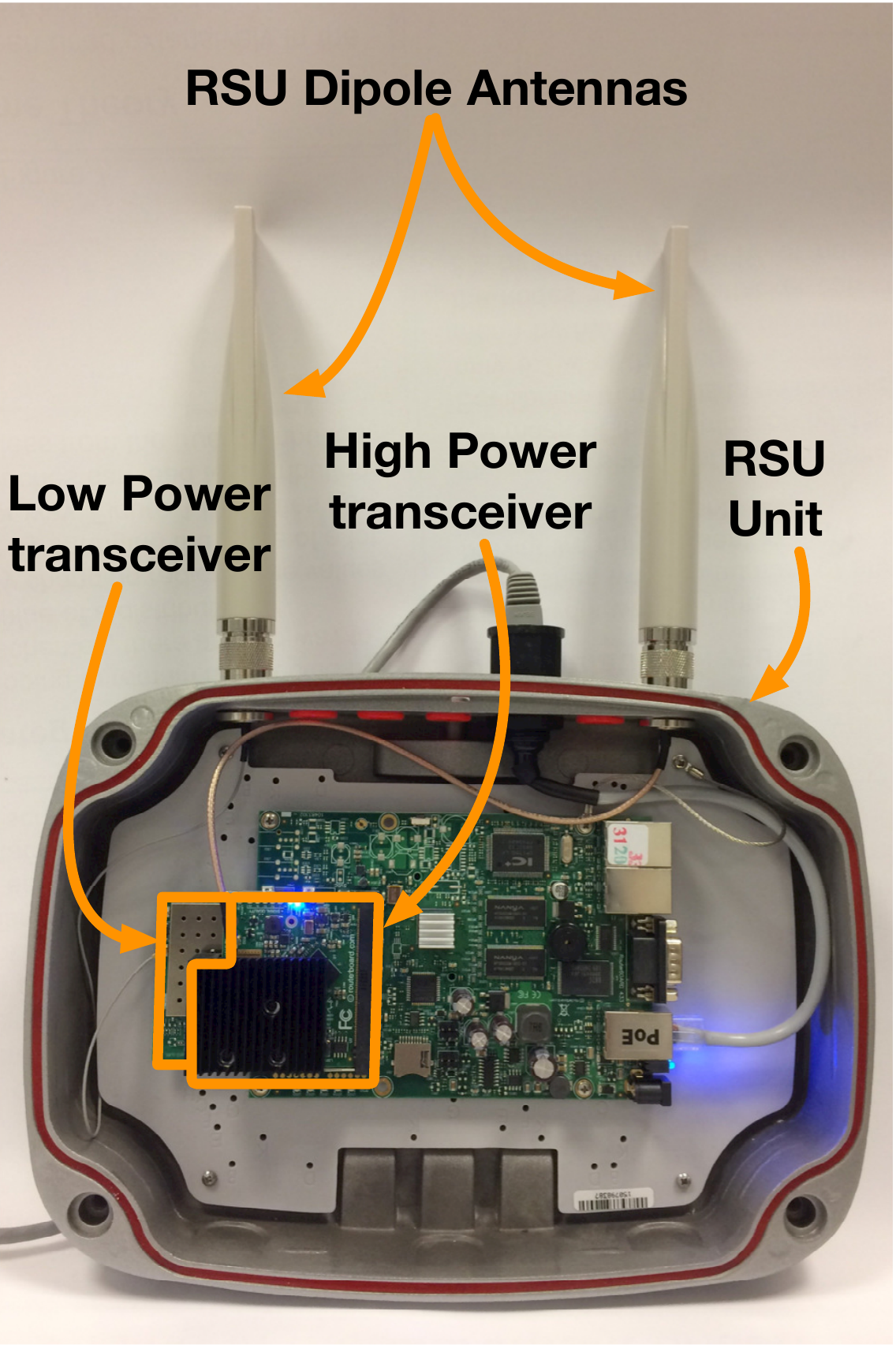}}
\end{minipage}%
\begin{minipage}{.69\linewidth}
\centering
\subfloat[OBU antenna mounted on the roof of the car~\cite{adHocNowCityScale}.]{\label{fig:car}\includegraphics[width=0.98\linewidth]{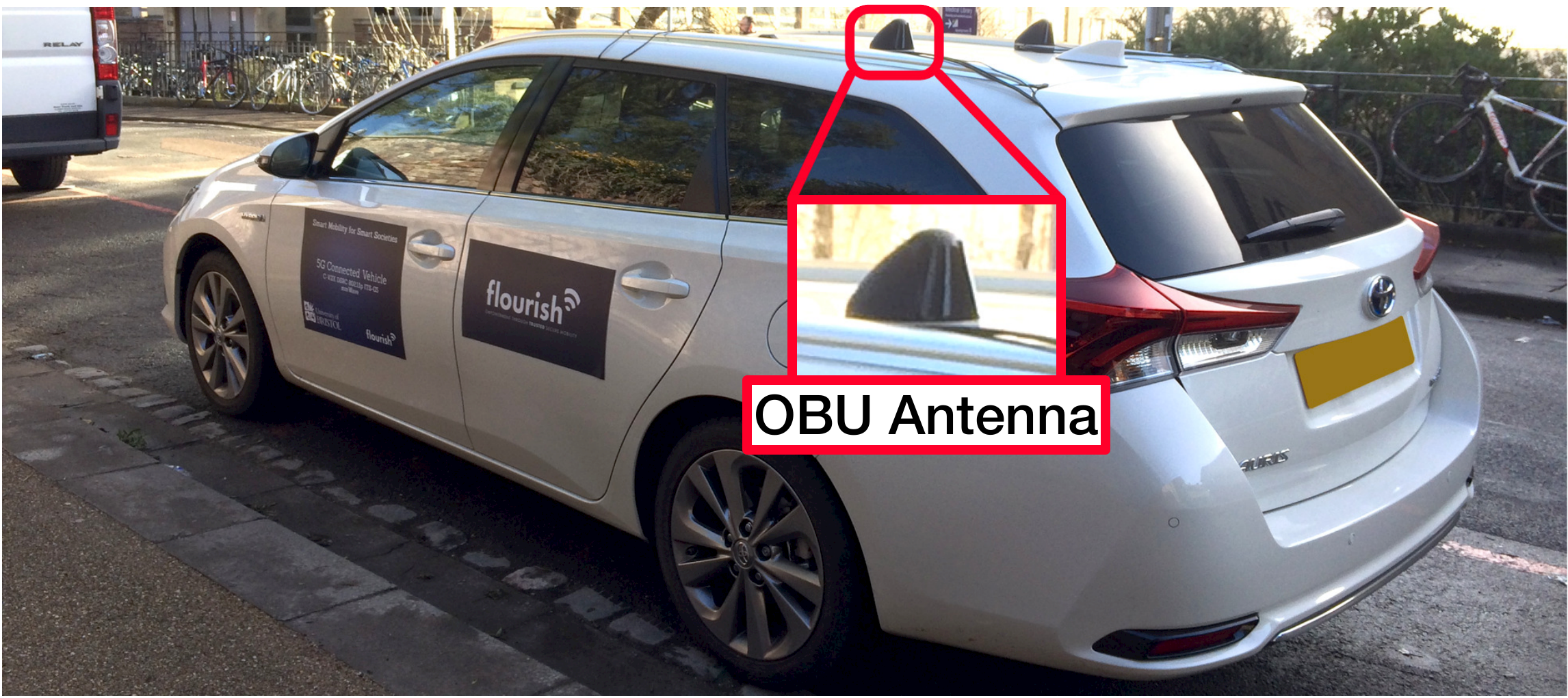}}
\end{minipage}\par\medskip
\caption{Our prototyped ITS-G5 testbed. We designed both RSUs and OBUs units, equipped them with different antennas and conducted our trials around Bristol, UK.}
\label{fig:unitsAll}
\end{figure*}

For our experimentation, we prototyped
an open-source IEEE 802.11p/DSRC testbed (Fig.~\ref{fig:unitsAll}). Our testbed consists of four different components. The core of our system is our vehicular communication nodes (Figs.~\ref{fig:close} and~\ref{fig:rsu}). These devices are responsible for generating, transmitting and receiving all the ITS-G5 network interactions between the vehicles and the road-side infrastructure. Sec.~\ref{sub:vehComm} will present both devices with further details.

Our communication devices record all the ITS-G5 interactions onto a data storage unit. As for the RSUs, this is a server connected to the network of the University of Bristol. In the case of the OBUs, the data storage facility is provided by a Raspberry Pi connected to a Solid-State Drive (SSD) hard drive, and then interfaced to the communication node by means of a \SI{100}{\mega\bit} Ethernet link (Fig.~\ref{fig:close}).

The onboard clock pertaining to each device are usually prone to drift, typically because of the inexpensive crystal oscillator circuits they employ. Moreover, different devices may exhibit vastly different behaviors when subject to vibration or shocks.
In order to reduce the drifting on each clock, we connected both our communication nodes to a Network Time Protocol (NTP) unit. We chose LeoNTP~\cite{ntpServer} as our NTP server solution, which is a Stratum 1 NTP server synchronized via Global Positioning System (GPS). That is, we ensured a high degree of synchronization between the devices. Also, to account for the drifting due to vibrations or shocks (especially for the mobile devices), we forced each device to periodically re-synchronize its clock with the NTP server (every \SI{30}{\second}).

Finally, our system was equipped with a Global Navigation Satellite System (GNSS) GPS receiver~\cite{gpsReceiver}. The GPS receiver has been used to both fulfill the time synchronization task and providing accurate positioning information. In the next section, we will present how we encapsulated the position and the timestamps within each ITS-G5 message. Detailed specifications of the NTP server and the GPS receiver can be found in Table~\ref{table:gpsNTP}.

\begin{table}[t]
\renewcommand{\arraystretch}{1.06}
\centering
\caption{NTP server and GPS receiver specifications.}
\begin{tabular}{*{1}{p{.34\columnwidth}}*{1}{P{.44\columnwidth}}}
\multicolumn{2}{c}{NTP Server~\cite{ntpServer}}  \\ \hline \hline
\raggedleft NTP Accuracy & $<\SI{1}{\micro\second}$ \\
\raggedleft NTP Requests/sec & $>100.000$ \\
\raggedleft Stratum Level & Stratum 1 \\ \hline
\\[1mm]
\multicolumn{2}{c}{GNSS GPS receiver~\cite{gpsReceiver}}  \\ \hline \hline
\raggedleft GPS Precision & $<\SI{1}{\meter}$ \\
\raggedleft Position Acquisition Time & $<\SI{32}{\second}$ \\
\raggedleft Oscillator Type & Crystal (Real-time clock support) \\
\raggedleft Update Rate & up to \SI{5}{\hertz} \\ \hline
\label{table:gpsNTP}
\end{tabular}
\end{table}

\vspace{-1mm}
\subsection{Vehicular Communication Infrastructure}\label{sub:vehComm}
Our vehicular communication nodes consist of the following three key components. A single-board computer and two wireless transceivers connected to their accompanied antennas. The single board computer provides the processing power needed to operate the ITS-G5 stack. For that, we adopted a Mikrotik RB433 RouterBoard (with CPU \SI{300}{\mega\hertz}, \SI{64}{\mega\byte} RAM, \SI{64}{\mega\byte} storage space, x3 Ethernet slots, x3 MiniPCI slots)~\cite{rb433}. As for the transceivers, we employed two types of mini-PCI Network Interface Controllers (NICs) that were responsible for the exchange of the ITS-G5 Cooperative Awareness Messages (CAMs). Both the NICs have been operated by means of customized Linux Kernel drivers making the NICs IEEE 802.11p/DSRC-compliant. The first NIC model was a Mikrotik R52H~\cite{r52h} operating at \SI{25}{\dBm}, and regarded as the low-power (LP) NIC for the remaining of the paper. The second NIC, regarded as the high-power (HP) transceiver, was a Mikrotik R5SHPn~\cite{R5SHPn} with \SI{29}{\dBm} maximum transmission power. These two wireless interfaces were installed on both the RSUs and the OBUs.

We adopted the system design for both RSUs or OBUs, with the only difference between them being the adopted antennas. For each RSU, antennas are bolted onto the device while each OBU, being secured in the boot of each vehicle (Fig.~\ref{fig:boot}), is connected to an antenna set magnetically attached on its rooftop (Fig.~\ref{fig:car}). In particular, as for the RSU, each NIC was connected to a dipole antenna with a gain of \SI{7}{\dBi}. On the other hand, each OBU NIC was connected to a dipole antenna with a gain of \SI{5}{\dBi}. Our RSU devices were powered up via Power-over-Ethernet (PoE), while a battery pack was used for the OBUs to avoid fast fluctuations in current as a result of using inverters connected to lighter-style sockets. For simplicity, we will refer to the different transceivers and devices by means of a combination of their abbreviations (for e.g., LP-RSU will stand for the low power transceiver of an RSU). The specifications of each device used in our trials are summarized in Table~\ref{table:characteristics}.

\begin{table*}[t]
\renewcommand{\arraystretch}{1.06}
\centering
    \caption{Wireless network interface controllers Specifications.}
    \label{table:characteristics}
    \begin{tabular}[b]{r|C|C|C|C}
                 &  LP-RSU             & LP-OBU             &  HP-RSU & HP-OBU                             \\ \hline \hline
    Model        &  \multicolumn{2}{c|}{Mikrotik R52H~\cite{r52h}}      &  \multicolumn{2}{c}{Mikrotik R5SHPn~\cite{R5SHPn}}         \\ \hline
    Chipset        &  \multicolumn{2}{c|}{AR5414}      &  \multicolumn{2}{c}{AR9220}         \\ \hline
    Linux Driver &  \multicolumn{2}{c|}{ath5k}              &  \multicolumn{2}{c}{ath9k}                   \\ \hline
    \multirow{2}{*}{Operational Frequency} &  \multicolumn{2}{c|}{\SIrange{2.192}{2.539}{\GHz}} &  \multicolumn{2}{c}{\multirow{2}{*}{\SIrange{4.800}{6.075}{\GHz}}}    \\
                                           &  \multicolumn{2}{c|}{\SIrange{4.920}{6.100}{\GHz}} &   \\ \hline
    TX Power     &
    \multicolumn{2}{c|}{\SI{25}{\dBm}}      &  \multicolumn{2}{c}{\SI{29}{\dBm}}           \\ \hline
    Antenna Gain &  \SI{7}{\dBi}  & \SI{5}{\dBi}  &  \SI{7}{\dBi}   &     \SI{5}{\dBi} \\ \hline
    Channel Bandwidth    &  \multicolumn{4}{c}{\SI{10}{\mega\hertz}}                                              \\ \hline
    $CW_{\mathrm{min}}, CW_{\mathrm{max}}$ &  \multicolumn{4}{c}{$\left[15, 1023\right]$} \\ \hline
    MCS    &  \multicolumn{4}{c}{QPSK \nicefrac{1}{2}} \\ \hline
    \end{tabular}
\end{table*}

\subsection{Vehicular Communication Operative System}
\begin{figure}[t]
\centering
\includegraphics[width=1\columnwidth]{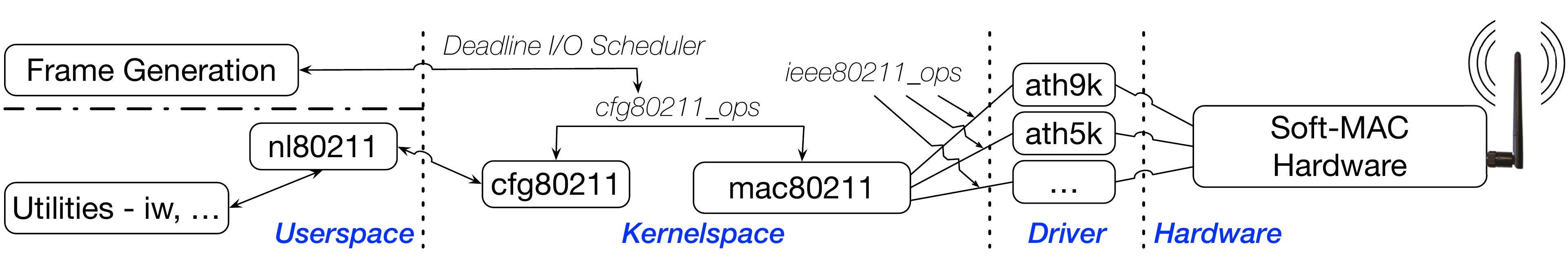}
    \vspace{-5mm}
    \caption{Linux Kernel Modules modified to enable the IEEE 802.11p/DSRC capabilities in our system~\cite{adHocNowCityScale}.}
    \label{fig:drivers}
\end{figure}

The operating system chosen for our devices was a low-latency OpenWRT Linux distribution\footnote{OpenWRT Barrier Breaker Release no. 14.07 - https://openwrt.org/}. Each transceiver operates on top of a different Atheros chipset and requires a different driver. The HP transceivers, operating via the AR9220 chipset, require the ath9k Linux driver while the LP ones, hosting an AR5414 chipset, require the ath5k Linux driver. Both were modified accordingly to enable IEEE 802.11p/DSRC functionalities~\cite{adHocNowCityScale}. The Linux kernel modules that we modified have been summarized in Fig.~\ref{fig:drivers}.

The software modules \emph{cfg80211} and \emph{nl80211} bridge the user and kernel space and offer the utility functionalities associated with 802.11. The \emph{mac80211} subsystem is the general driver framework and allows finer control of the hardware. The \emph{iw} tool is used for configuring the utility of the NIC and is based on the \emph{nl80211} netlink interface. Furthermore, \emph{cfg80211\_ops} and \emph{ieee80211\_ops} define the operations and the callbacks between the different blocks. The Outside the Context of a BSS (OCB) mode was enabled in the MAC layer, allowing all NICs within range to communicate, without being authenticated/associated. Besides, the OCB mode commands were added in a modified version of the \emph{iw} utility to enable the functionality mentioned above. The values for the contention windows and the MCSs were chosen to follow the regulations for the ITS-G5 standard specifications.

\subsection{IEEE 802.11p/DSRC CAMs and Logging Interfaces}\label{sub:camLogging}

In our system, IEEE 802.11p/DSRC CAMs are exchanged between all the devices. All CAMs are generated in the Facilities layer of the ITS-G5 stack~\cite{etsiCam}. The Facilities layer is responsible for handling all the packets generated from the different ITS application, such as the CAMs or the  Decentralized Environmental Messages (DENMs). Then, the Facilities layer forwards all the generated messages to the lower protocol layers that are responsible for the actual transmission.

In our testbed, we built on top of a pre-existing ITS-G5 stack~\cite{adHocNowCityScale} and we implemented a beaconing interface in charge of generating CAMs.
To log all these network interactions we designed two different logging interfaces. The first one records the exchanged CAMs as CSV-formatted files, while the second stores them as PCAP traces. The first format enabled us to easily reconcile the sequences of received and transmitted CAMs -- thus enabling us to quickly extract preliminary communication metrics. Then, PCAP traces enable experts to investigate further key performance indicators not considered in this paper and not already included in the CSV-formatted part of our dataset.

\begin{table*}[t]
\renewcommand{\arraystretch}{1.06}
\centering
    \caption{Name and description of the fields found in the logged CAM messages, both at the TX and RX sides.}
    \label{table:fields}
    \begin{tabular}[b]{rP{7.6cm}|rP{5.1cm}}
    \textbf{TX Fields} & \textbf{Description} & \textbf{RX Fields} & \textbf{Description} \\ \hline \hline
    TX-REQ-CAM & Type of packet & RxMAC & MAC address of the CAM source node \\
    Protocol & Dummy field (is always 1)~\cite{etsiCam}  & RX-REQ-CAM &  Type of packet \\
    StationID & Dummy field with random value~\cite{etsiCam} & Validation & Show if constraints failed in ASN.1~\cite{etsiCam} \\
    GenDeltaTime & The remainder of $\mathrm{Timestamp}/65546$~\cite{etsiCam} & Protocol & Dummy field (is always 1)~\cite{etsiCam} \\
    SeqNum & Sequence number of transmitted packet & StationID & The StationID of the transmitter~\cite{etsiCam} \\
    GpsLon & Current longitude of the node & GenDeltaTime & The remainder of $\mathrm{Timestamp}/65546$ at the moment of transmission~\cite{etsiCam} \\
    GpsLat & Current latitude of the node & SeqNum & Sequence number of received packet \\
    CamLon & Quantized value of \emph{GpsLon} -- the value encapsulated in a CAM & GpsLon & Current longitude of the node \\
    CamLat & Quantized value of \emph{GpsLat} -- the value encapsulated in a CAM & GpsLat & Current latitude of the node \\
    GpsSpeed & Current speed of the node & GpsSpeed & Current speed of the node \\
    CamSpeed & Quantized value of \emph{GpsSpeed} -- the value encapsulated in a CAM & SpeedConf & The way that speed is encoded~\cite{etsiCam} \\
    Timestamp & Timestamp when CAM is generated (in epoch time) & VehHeading & GPS-acquired heading of the node \\
    CamLength & The length of the transmitted CAM & CamLon & Encapsulated longitude value in a CAM  \\
    & & CamLat & Encapsulated latitude value in a CAM \\
    & & CamSpeed & Encapsulated speed value in a CAM \\
    & & Timestamp & Timestamp when a CAM is received \\ \hline

    \end{tabular}
\end{table*}

Considering the CSV-formatted part of our dataset, the data are organized in a tabular format with their fields summarized in Table~\ref{table:fields}. In particular, for what concerns the transmitted CAMs (namely, TX side), the position acquired from the GPS is represented as \emph{GpsLon} and \emph{GpsLat}, these being the longitude and latitude values of the transmitter, respectively. The fields \emph{CamLon} and \emph{CamLat} are the quantized values of the GPS that are encapsulated in a CAM.
The sequence number of the generated CAMs is shown in the \emph{SeqNum} field (starting from zero when each device boots up). This value can be later used to correlate the list of transmitted and received CAMs. The \emph{GpsSpeed} and \emph{CamSpeed} represent the speed of the vehicle, with the first being the acquired value from the GPS and the latter the quantized value encapsulated in the CAM. The \emph{Timestamp} is the time that the packet is generated, given in Unix Epoch format. Finally, the \emph{CamLength} is the length of the transmitted CAM in bytes.

For what concerns the received CAMs (namely, RX side), the most important fields are the following. The \emph{RxMAC} is the source MAC address. \emph{CamLon} and \emph{CamLat} are the source node position coordinates encapsulated in the received packet. The \emph{GpsLon} and \emph{GpsLat} values represent the current longitude and latitude of the receiver, acquired from the GPS receiver and quantized later. The \emph{Timestamp} field is the time that a CAM is received, given in Unix Epoch format and \emph{SeqNum} is the encapsulated sequence number of the CAM received.

On both sides, the fields \emph{TX-REQ-CAM} and \emph{RX-REQ-CAM}, are used to tell if a specific log entry refers to a transmitted or received CAM. Finally, fields such as \emph{Protocol}, \emph{Validation}, \emph{GenDeltaTime}, \emph{VehHeading} and \emph{StationID} have not been considered in our performance investigation.

As for the PCAP traces, they include all the exchanged network interactions. PCAP traces, being the industrial and scientific format for raw-data recording for network applications,  are compatible with a wide variety of programs and provides direct access to the binary-level of the network exchanges. In addition, every received message is recorded along with its RSSI value, which is one of the key metrics that will be considered in our performance investigation.
The entire database with our exchanged network interactions can be found in~\cite{dataset}. More details about the way these data are stored or could be potentially used in the future can be found in~\cite{dataInBrief}. Finally, a repository with all the scripts used to parse and process the results can be found in~\cite{repository}.

\begin{figure}[t]
\centering
\includegraphics[width=1\columnwidth]{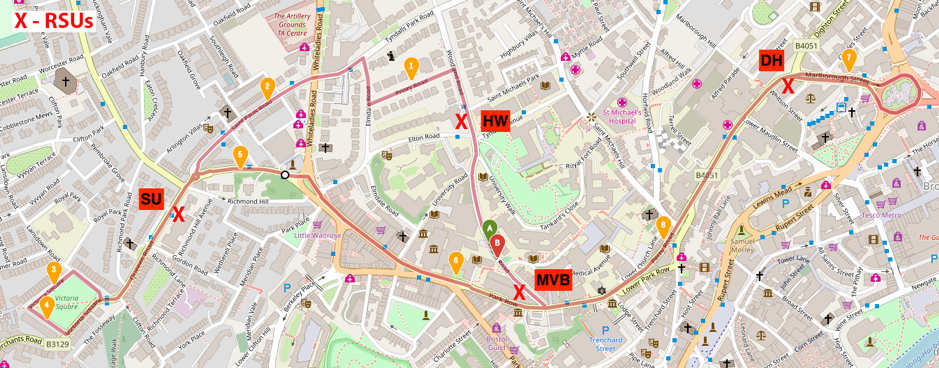}
    \caption{The routes of the vehicles and the positions of the four RSUs.}
    \label{fig:rsuPositions}
\end{figure}

\section{Vehicular Trials}\label{sec.3}

For our experimental evaluation, we deployed four RSUs around the campus of the University of Bristol. The positions of the RSUs are shown in Fig.~\ref{fig:rsuPositions} (marked with the red X’s). To maximize the heterogeneity in terms of the road and building layout and investigate how these factors affect the network performance, we positioned our RSUs in the four positions shown on the map with the following characteristics:
\begin{itemize}
    \item \emph{Merchant Ventures Building (MVB)}: The RSU is mounted at a balcony of the building, close to a blind T-junction, on a curvy road at a height of \textasciitilde\SI{8}{\meter}.
    \item \emph{Dorothy Hodgkin (DH) building}: Curvy wide road, one of the main arteries of the City of Bristol, UK. The RSU is mounted at the balcony at \textasciitilde\SI{12}{\meter}, providing good Line-of-Sight (LOS) coverage on the curved road.
    \item \emph{Hawthorns (HW) building}: Straight road with light foliage on one side of the RSU. The RSU is mounted on a wall of a building at \textasciitilde\SI{5}{\meter}.
    \item \emph{Students Union (SU) building}: Straight road, RSU mounted at a balcony at \textasciitilde\SI{25}{\meter}. This site provides full coverage on the road in front of the building and between the two roundabouts.
\end{itemize}
For simplicity, in the rest of this paper, we will refer to all the RSUs with their abbreviation.

During our trials, two vehicles have been used. Both the vehicles followed the same route that is shown in Fig.~\ref{fig:rsuPositions}. One vehicle was driving in a clockwise direction, while the other one was driving anti-clockwise. Each vehicle was equipped with an OBU as discussed before.

During the four days of trials, we measured the performance of the system over different frequencies. More specifically, during the first day, we tested the performance over the DSRC frequency band. The findings from this day were later compared against the measurements from days 2-4, where we operated our system over three different ISM-2.4 and ISM-5 bands. Details about the center frequency where our system has been operated is summarized in Table~\ref{table:freq}.

\begin{table}[t]\vspace*{-5mm}
\renewcommand{\arraystretch}{1.06}
\centering
\caption{The frequencies used throughout the four days of trials.}
    \begin{tabular}{r|M{2cm}|M{2cm}|M{1cm}}
    \textbf{Day} & \textbf{HP Transceiver} & \textbf{LP Transceiver} & \textbf{Band} \\ \hline \hline
    Day 1 & \SI{5.9}{\GHz} & \SI{5.89}{\GHz} & DSRC \\
    Day 2 & \SI{5.2}{\GHz} & \SI{2.437}{\GHz} & ISM \\
    Day 3 & \SI{5.18}{\GHz} & \SI{2.412}{\GHz} & ISM \\
    Day 4 & \SI{5.32}{\GHz} & \SI{2.462}{\GHz} & ISM \\ \hline
    \end{tabular}\vspace*{-5mm}
\label{table:freq}
\end{table}

\section{Performance Investigation}\label{sec.4}

\begin{figure*}[t]
\minipage{0.49\textwidth}
\centering
    \includegraphics[width=1\columnwidth]{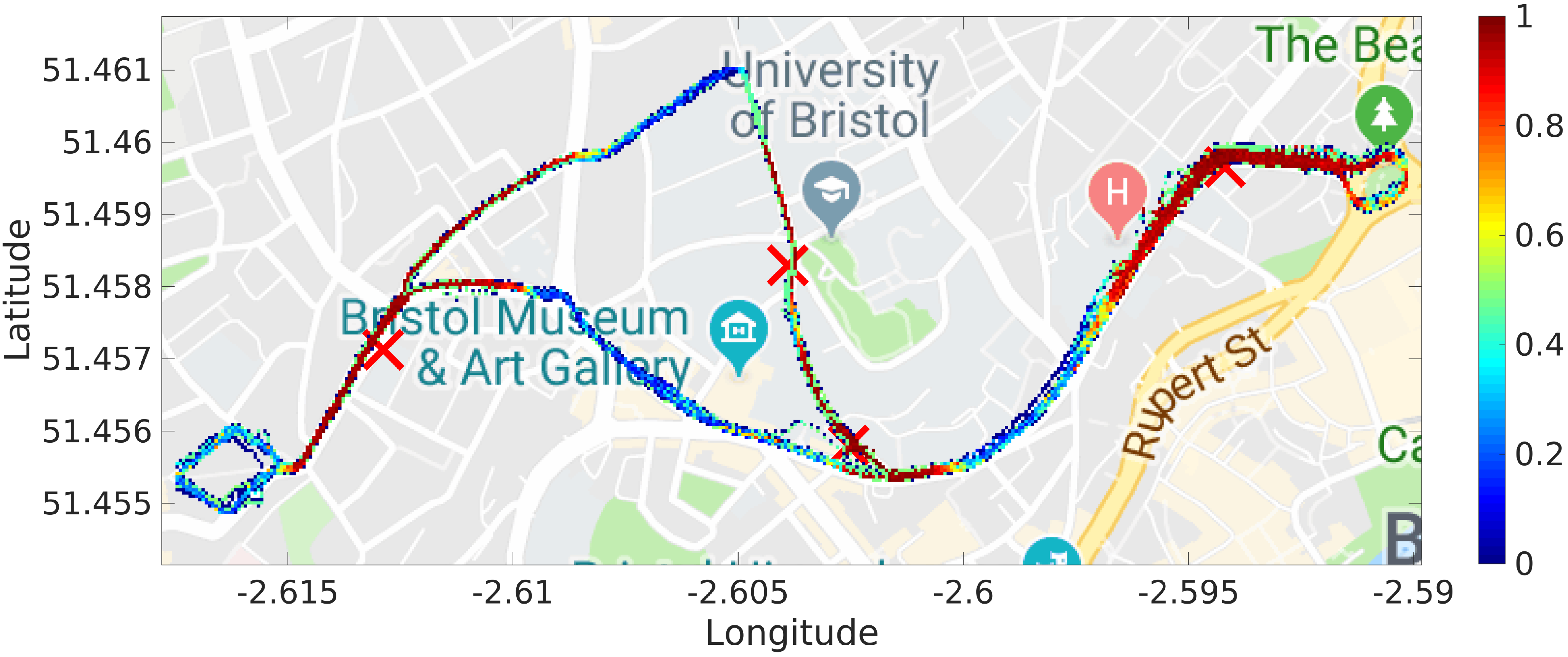}
    \vspace{-5mm}
    \caption{Heatmap results for the DSRC band. This figure shows the PDR for the first day of our experimental trials and the HP transceiver.}
    \label{fig:heatmap}
\endminipage\hfill
\minipage{0.49\textwidth}
\centering
\centering
    \includegraphics[width=1\columnwidth]{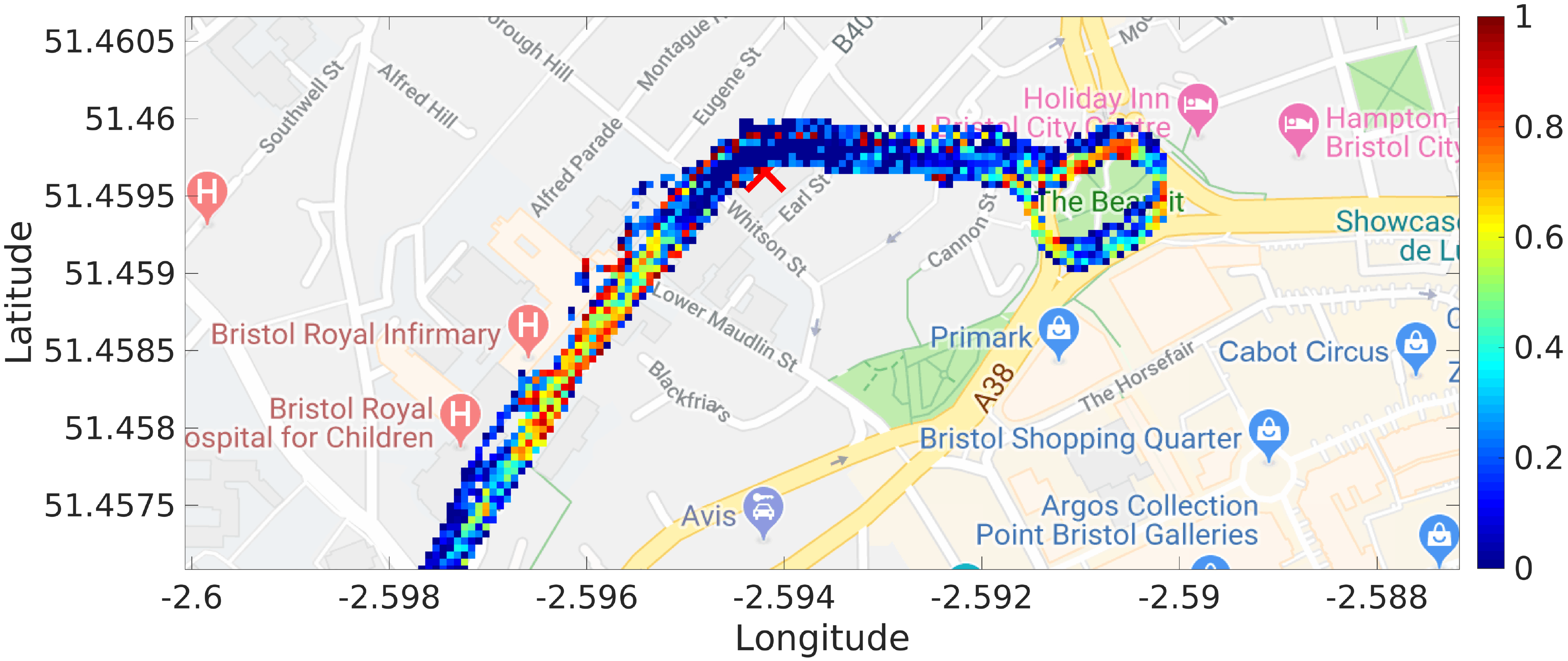}
    \vspace{-5mm}
    \caption{The MSE difference, represented as a heatmap, for the tiles around the DH-RSU. This is a comparison between days 1 and 3.}
    \label{fig:heatmapMean}
\endminipage\hfill
\end{figure*}

Various Key Performance Indicators (KPIs) were used to observe the impact of operating an ITS-G5 system over ISM bands, from the perspective of both the OBUs and RSUs. As discussed in Sec.~\ref{sec.3}, our experimental trials lasted for four days. Every day, we conducted our trials during two different sessions (morning/afternoon). All the results presented in this section, will be the average of each day, namely, the average of both the morning and the afternoon session.

All the devices in our system generated and transmitted one ITS-G5 CAM per NIC every \SI{10}{\milli\second}. Each CAM was then logged in a two-way fashion, as discussed in Sec.~\ref{sub:camLogging}. The overall CAM size \emph{CamLength} was always 109 bytes.
During our field trials, there was no provision for cyber-security related features.
Each day a different frequency was chosen (Table~\ref{table:freq}). The LP transceiver was operated at the ISM-2.4 frequency band while the HP one was operated at the ISM-5.

\begin{figure*}[t]
\minipage{0.49\textwidth}
\centering
    \includegraphics[width=1\columnwidth]{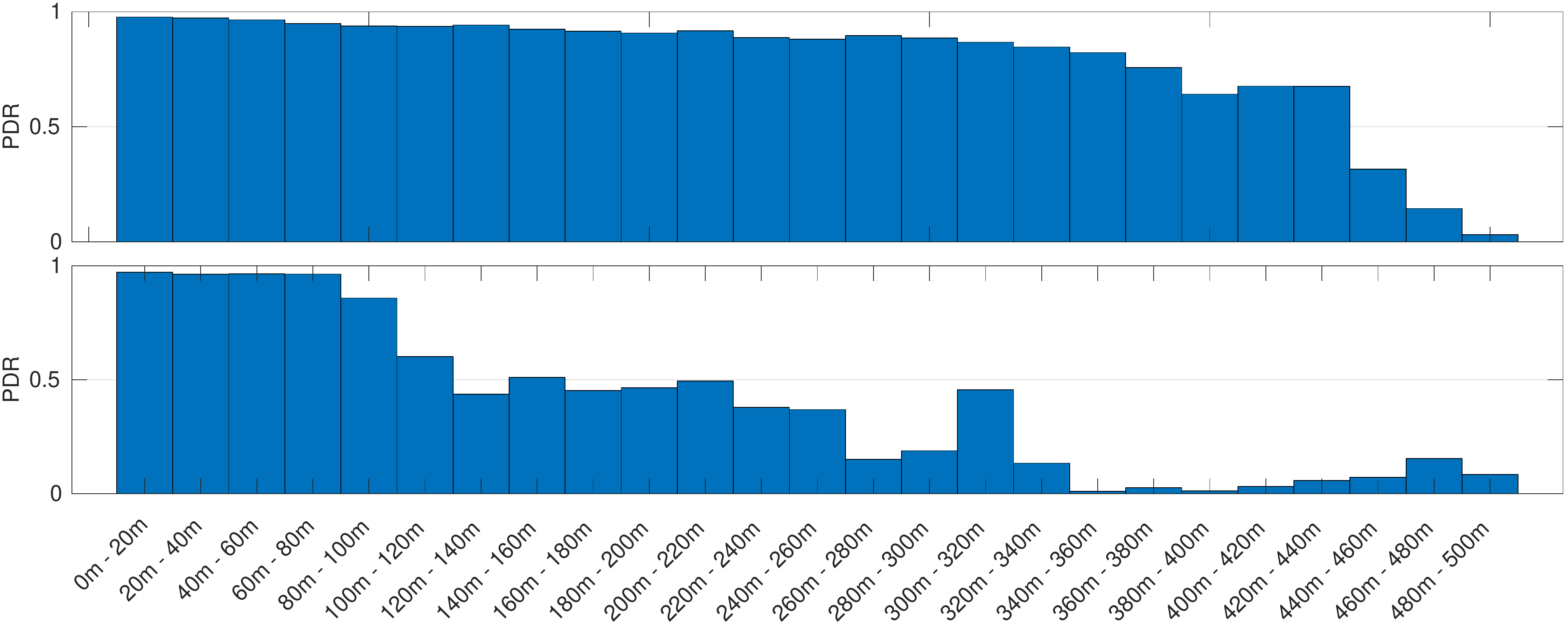}
    \vspace{-6mm}
    \caption{The awareness horizon for the DSRC band and the HP transceiver of the first vehicle. Top plot: DH-RSU, bottom plot: HW-RSU.}
    \label{fig:awarenessHorizonDay1}
\endminipage\hfill
\minipage{0.49\textwidth}
\centering
\centering
    \includegraphics[width=1\columnwidth]{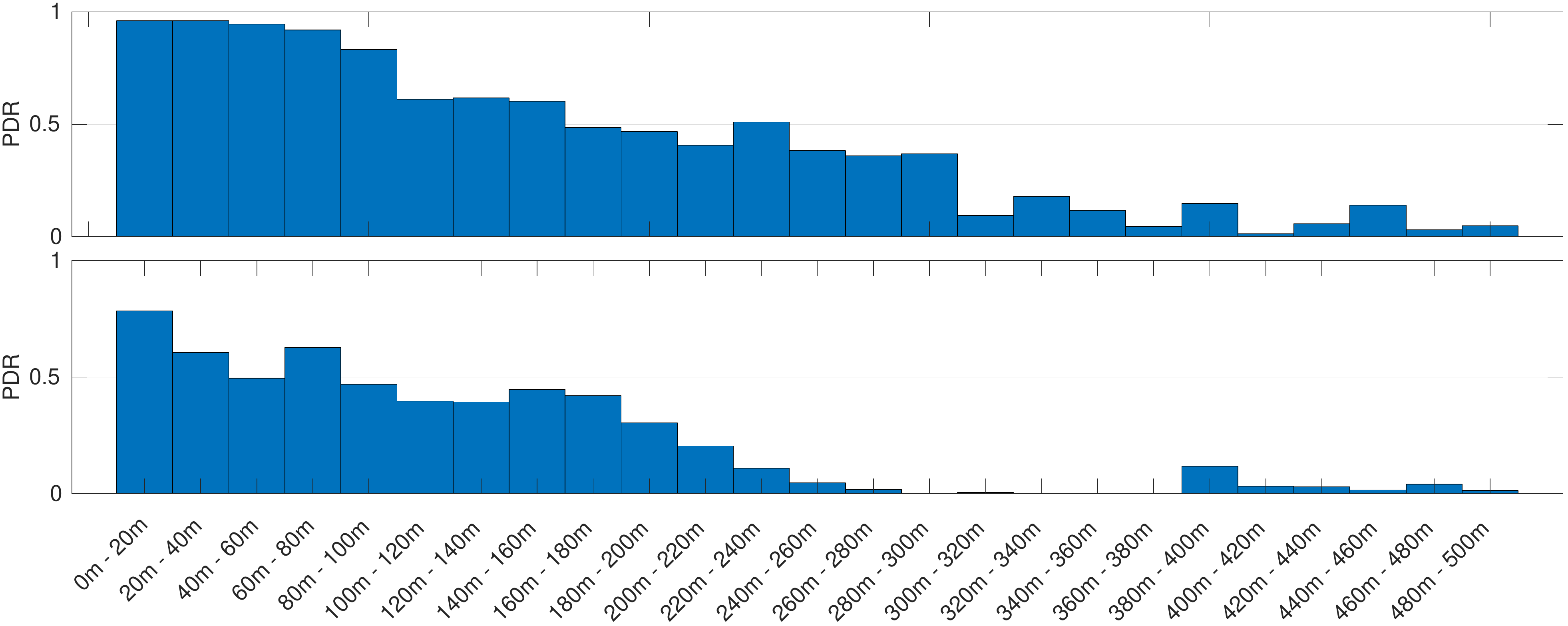}
    \vspace{-6mm}
    \caption{The awareness horizon for the third day and the HP transceiver of the first vehicle. Top plot: DH-RSU, bottom plot: HW-RSU.}
    \label{fig:awarenessHorizonDay3}
\endminipage\hfill

\vspace{3mm}

\minipage{0.49\textwidth}
\centering
    \includegraphics[width=1\columnwidth]{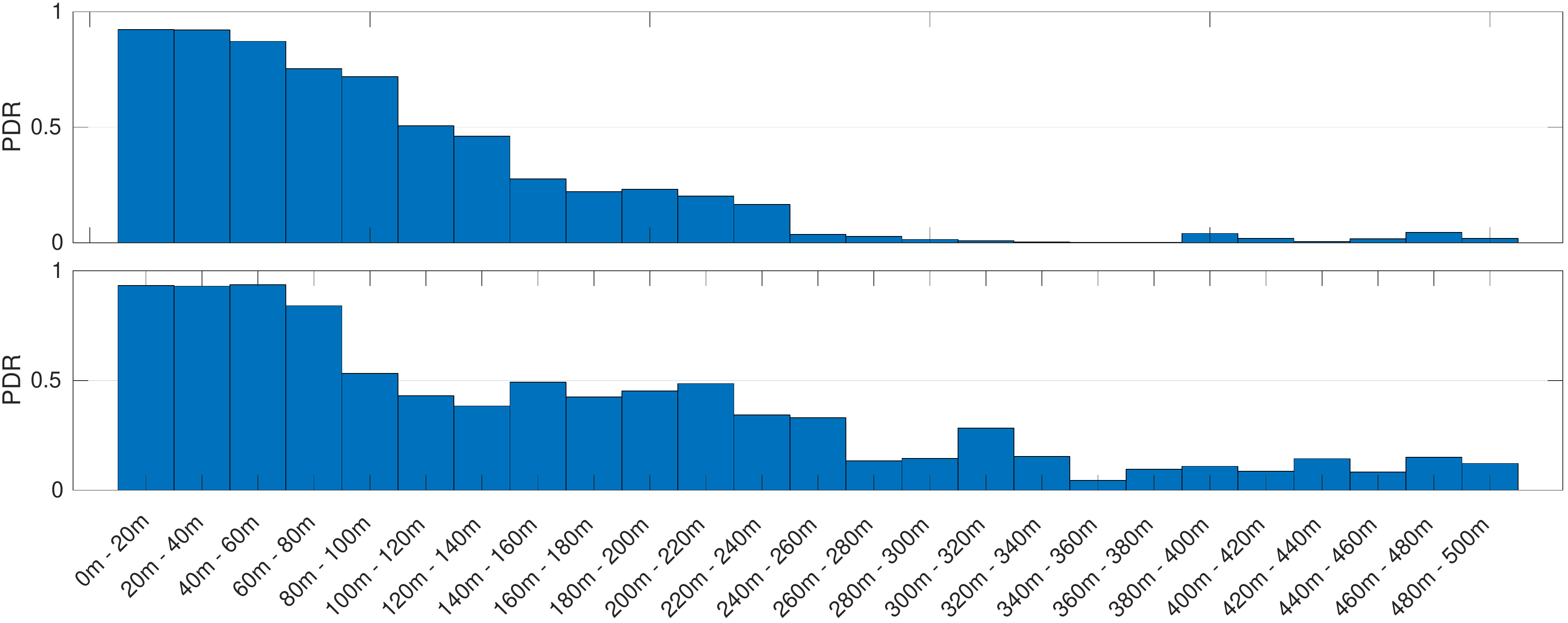}
    \vspace{-6mm}
    \caption{The awareness horizon for the DSRC scenario and the LP transceiver of the first vehicle. Top plot: DH-RSU, bottom plot: HW-RSU.}
    \label{fig:awarenessHorizonDay1LP}
\endminipage\hfill
\minipage{0.49\textwidth}
\centering
\centering
    \includegraphics[width=1\columnwidth]{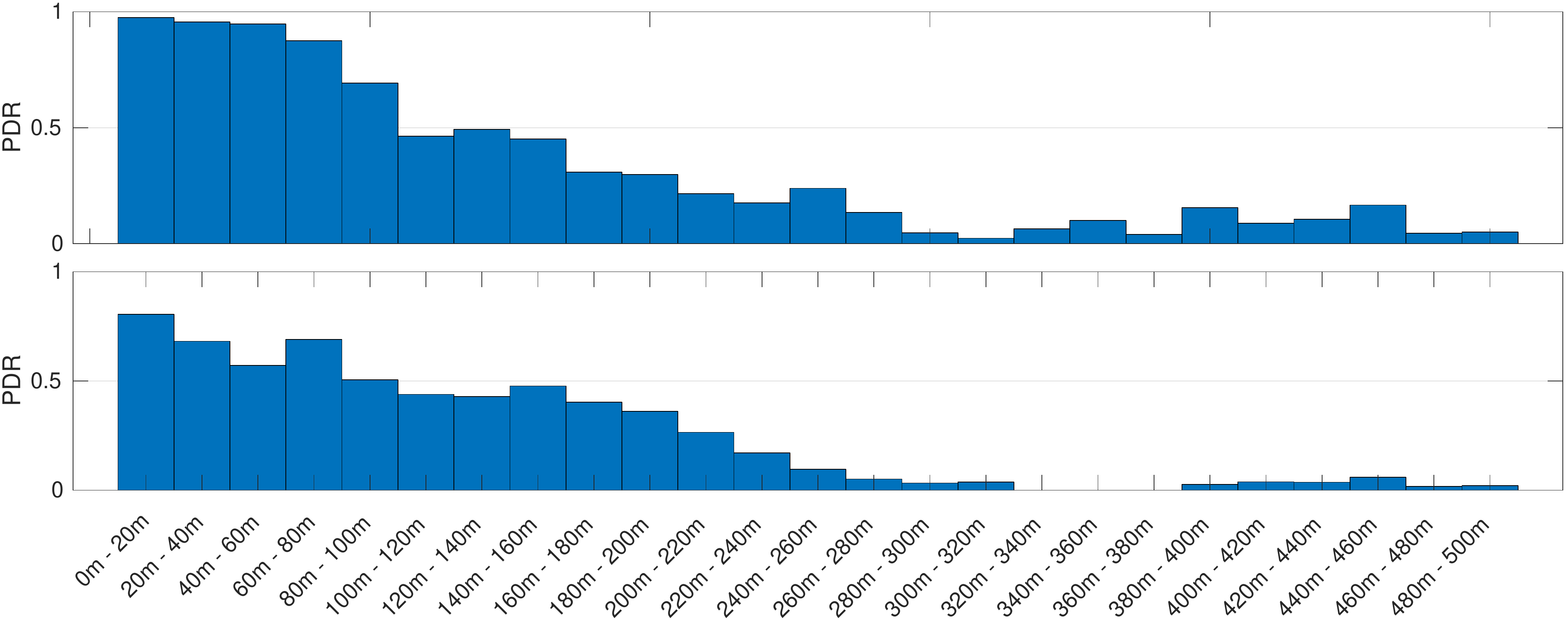}
    \vspace{-6mm}
    \caption{The awareness horizon for the third day and the LP transceiver of the first vehicle. Top plot: DH-RSU, bottom plot: HW-RSU.}
    \label{fig:awarenessHorizonDay3LP}
\endminipage\hfill
\end{figure*}

\begin{table}[t]\vspace*{-5mm}
\renewcommand{\arraystretch}{1.06}
\centering
\caption{MAD and MSE comparison between the ISM and DSRC bands}
    \begin{tabular}{r|M{1.5cm}|M{1.5cm}|M{1.2cm}|M{1.2cm}}
    \textbf{Day} & \textbf{MAD HP} & \textbf{MAD LP} & \textbf{MSE HP} & \textbf{MSE LP} \\ \hline \hline
    Day 2 & $0.2246$ & $0.1473$ & $0.1048$ & $0.0496$ \\
    Day 3 & $0.2978$ & $0.1525$ & $0.1676$ & $0.0538$ \\
    Day 4 & $0.2257$ & $0.1552$ & $0.1082$ & $0.0559$ \\ \hline
    \end{tabular}\vspace*{-5mm}
\label{table:mseResults}
\end{table}

Fig.\ref{fig:heatmap} shows the PDR as it is perceived at the RSU-side. This particular figure refers to the experiments run in the DSRC band (day 1) and the HP transceivers. The displayed PDR is the average of all the packet delivered over the packets transmitted from each OBU, when driving within the boundaries of a map squared tile that are $\SI{5}{\meter}^2$ wide. As such, each tile covers roughly the width of a road lane. As expected, being the DH-RSU installed higher up on the building, it provides the best coverage amongs all the RSUs. DH-RSU is followed by the SU-RSU, MVB-RSU and finally by the HW-RSU providing the worst coverage range and PDR. In addition, compared to the HP tranceiver, the performance of the LP transceiver is on average $25\%$ lower for the first day.

Table~\ref{table:mseResults} presents the Mean-Squared Error (MSE) and the Mean Absolute Difference (MAD) of the PDR per-tile. This table compares the results from the first day of experiments (DSRC experiment) against the results from the remaining days (ISM experiments). In particular, for the LP transceivers, we observe that both KPIs are comparable for all days, showing marginal variations. On the other hand, in the case of HP transceivers, we observe a more significant alteration during the third day, which can be caused by a different level of interference in the considered ISM frequency (namely, the IEEE 802.11a channel 36).

Fig.~\ref{fig:heatmapMean} also shows the MSE per tile, for the area around the DH-RSU and the HP transceiver. This figure is the comparison of the ISM-5 band with the DSRC one, for the worst case scenario we observed before (Table~\ref{table:mseResults}). The MSE significantly increases as we move away from the DH-RSU. The increased interference from the surrounding WiFi networks and the attenuated signal because of the RSU-OBU distance increases the packet loss. A similar behavior can be observed around the surrounding areas of the remaining RSUs and for all days.
The above behaviour shows that on both frequency bands, the PDR performance for closer distances is comparable. Of course, for longer distances impact of the interference from the surrounding devices is more observable.

\begin{figure*}[t]
\minipage{0.49\textwidth}
\centering
    \includegraphics[width=1\columnwidth]{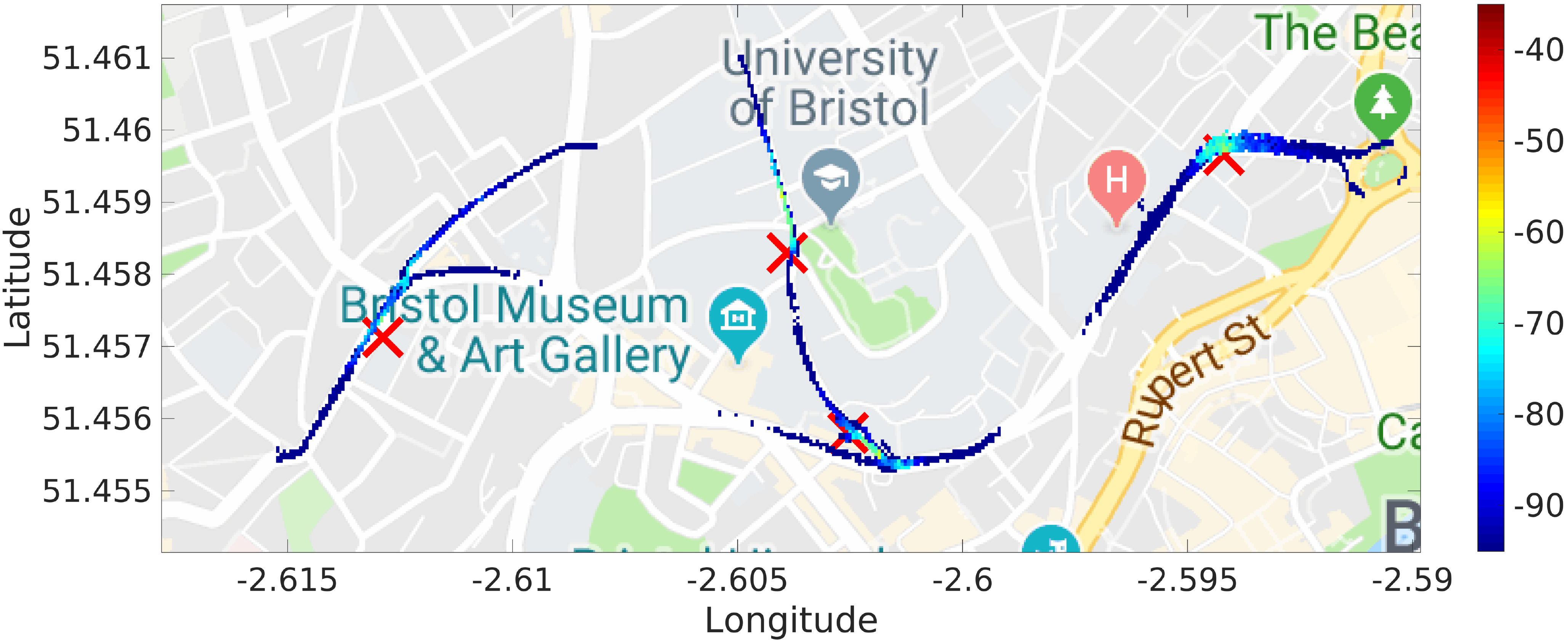}
    \vspace{-7mm}
    \caption{Heatmap RSSI results for the LP transceiver, during the first day.}
    \label{fig:rssiLPDay1}
\endminipage\hfill
\minipage{0.49\textwidth}
\centering
\centering
    \includegraphics[width=1\columnwidth]{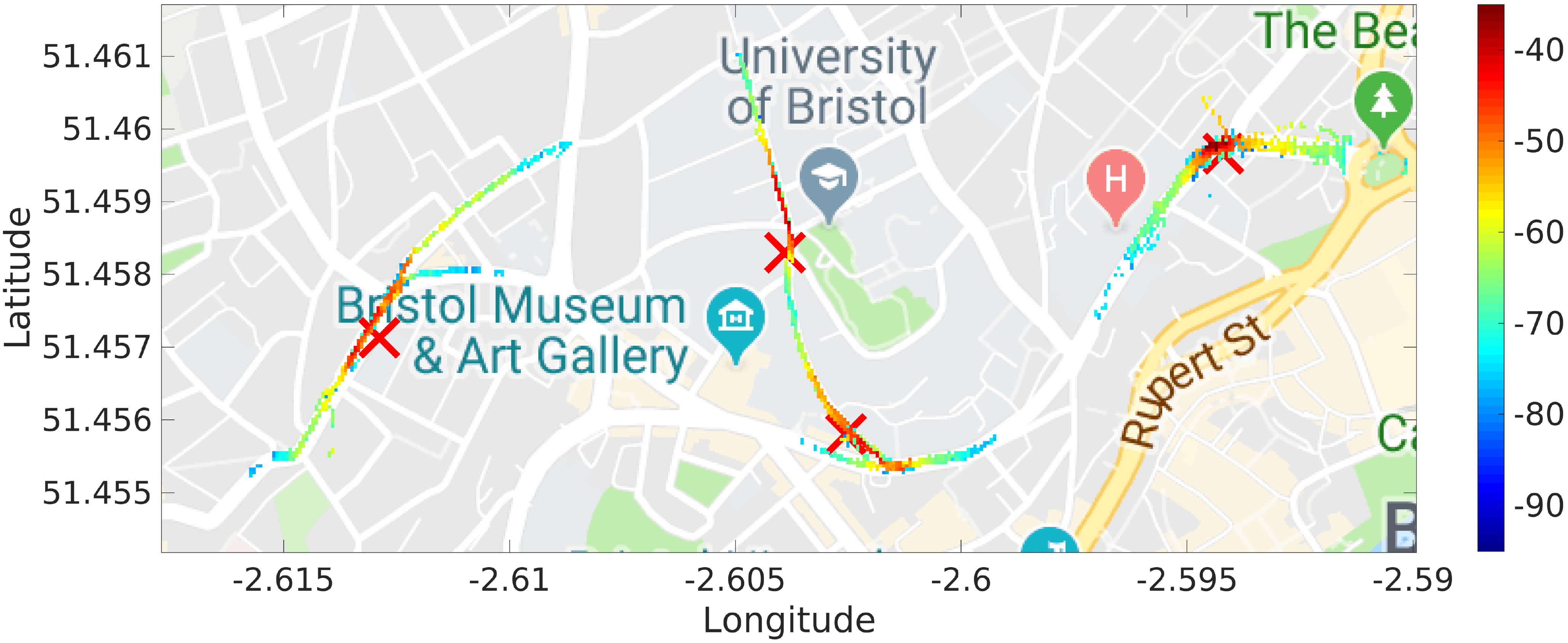}
    \vspace{-7mm}
    \caption{Heatmap RSSI results for the LP transceiver, during the third day.}
    \label{fig:rssiLPDay3}
\endminipage\hfill
\end{figure*}

Figs.~\ref{fig:awarenessHorizonDay1} --~\ref{fig:awarenessHorizonDay3LP} show the awareness horizon from the perspective of one vehicle. The awareness horizon is defined as the PDR perceived from an OBU as a function of the Euclidean distance to an RSU. Having the average PDR per distance interval, we can have an indication of the perceived awareness at the receiver side, based on the sensor data that the transmitter managed to successfully send. Fig.~\ref{fig:heatmap} and  Table~\ref{table:mseResults} show that the worst system behavior, compared a system operated onto the DSRC band, was on our third day of trials. The best RSU (in terms of PDR and coverage) was the DH-RSU, while the worst being the HW-RSU. Figs.~\ref{fig:awarenessHorizonDay1} and~\ref{fig:awarenessHorizonDay1LP} refer to the first day, for DH-RSU (top) and HW-RSU (bottom) and both transceivers. On the other hand, Figs.~\ref{fig:awarenessHorizonDay3} and~\ref{fig:awarenessHorizonDay3LP} are equivalent results referring to the third day.
As expected, the HP transceivers, due to their increased transmission power manage to achieve increased PDR for longer distances, compared to the LP ones. For the HP transceiver and a distances smaller than $\SI{80}{\meter}$, we see that the DH-RSU has comparable performance for all the different frequencies. However, for a distance greater than $\SI{80}{\meter}$, the impact of the interference is considerably worse. However, this is not the case for the LP transceiver. Even though the ISM-2.4 is heavily congested, we see that due to the lower operational frequency, we get comparable results on both days and RSUs. For distances smaller than $\SI{80}{\meter}$, HW-RSU performs slightly better in the DSRC frequency band. This behavior is believed to be due to the position of the RSUs. HW-RSU, located very close to various university's buildings and students accommodations, is heavily impaired by interferences. However, as shown, for longer distances, the decreased signal attenuation due to the frequency almost compensates with the effects of the existing interference. The decreased signal attenuation due to the frequency is confirmed in Figs.~\ref{fig:rssiLPDay1} and~\ref{fig:rssiLPDay3}. These figures show the perceived average RSSI for each map-tile. In particular, we observe that the RSSI of CAMs transmitted when the system is operated at the ISM-2.4 frequency band is significantly higher compared to the case when we refer to DSRC or ISM-5 bands, as expected.



\vspace{-2mm}
\section{Conclusions}\label{sec.5}
In this work we presented a large-scale performance investigation for the feasibility of operating an ITS-G5 system over unlicensed frequency bands. We conducted our performance evaluation using our prototyped real-world ITS-G5 testbed. Our experimental testbed was operated on both the licensed DSRC and the corresponding unlicensed ISM frequency bands. We compared the performance using different communication metrics, such as the PDR and the RSSI. From the presented results we showed that our system achieved comparable performance against DSRC, over the ISM-2.4 frequency band, while having an increased PDR of roughly $30\%$, when operated over the ISM-5 band. During the time of the trials, we were the only users over the DSRC band. Given the expected market penetration of CAVs, the licensed band results are expected to be worse in the future. Taking that into account, and given our experimental results we believe that operating an ITS-G5 system over the ISM frequency bands, and especially over the ISM-2.4 band, is a viable option for the future. To increase the reproducibility of the results, as well as pave the way for future research avenues (e.g. on deriving empirical models or exploiting cyber-security attacks), we recorded all the network interactions in a two-folded fashion and made our generated database publicly available.

\vspace{-2mm}
\section*{Acknowledgment}
This work is part of the FLOURISH Project, which is supported by Innovate UK, under Grant 102582.

\vspace{-2mm}
\bibliographystyle{IEEEtran}
\bibliography{IEEEabrv,papers}
\end{document}